\documentclass[prl,groupedaddress,reprint,superscriptaddress,floats,nofootinbib,preprintnumbers]{revtex4-1}
\usepackage{graphicx} 
\usepackage{float}
\usepackage{amsmath}
\usepackage{epstopdf}
\usepackage{xspace}
\usepackage{rotating}
\usepackage{longtable}
\usepackage{multirow}
\usepackage{booktabs}
\usepackage[breaklinks=true]{hyperref}
\hypersetup{
  colorlinks=true,
  linkcolor=blue,
  citecolor=blue,
  urlcolor=blue
}

\graphicspath{{./figs/}}
\usepackage{array,tabularx,epsfig,mathrsfs,graphicx,rotating}
\usepackage{ifthen}
\usepackage{amsfonts}
\usepackage{amssymb}

\newcommand{\be}{\begin{equation}}
\newcommand{\ee}{\end{equation}}
\newcommand{\bea}{\begin{eqnarray}}
\newcommand{\eea}{\end{eqnarray}}

\newcommand{\A}{\mathcal{A}}

\newcommand{\F}{\mathcal{F}}

\newcommand{\T}{\mathbf{T}}
\newcommand{\U}{\mathbf{U}}
\newcommand{\V}{\mathbf{V}}

\newcommand{\tr}{{\rm Tr}}
\renewcommand{\to}{\rightarrow}

\newcommand{\de}{\partial}

\newcommand{\beq}{\begin{equation}}
\newcommand{\eeq}{\end{equation}}
\renewcommand{\[}{\begin{equation}}
\renewcommand{\]}{\end{equation}}
\newcommand{\LL}{\mathscr{L}}

\newcommand{\DLR}{\mathbf{D}}

\newcounter{diagram}

\def\cF{{\cal F}}

\newcolumntype{C}[1]{>{\centering\let\newline\\\arraybackslash\hspace{0pt}}m{#1}}

\begin{document}

\begin{flushleft} 
IFT-UAM-CSIC-19-73  
\end{flushleft}
\begin{flushright} 
FTUAM-19-10 
\end{flushright}

%%%%%%%%%%%%%%%%%%%%%%%%%%%%%%%%%%%%%%%%%%%%%%%%%%%%%%%%%%%%%%%
\title{\boldmath
Non-Resonant Searches for Axion-Like Particles at the LHC 
}
%%%%%%%%%%%%%%%%%%%%%%%%%%%%%%%%%%%%%%%%%%%%%%%%%%%%%%%%%%%%%%%
\author{M. B. Gavela}
\email[]{belen.gavela@uam.es}
\affiliation{Departamento de Fisica Teorica, Universidad Autonoma de Madrid,
Cantoblanco, 28049, Madrid, Spain
}
\affiliation{Instituto de Fisica Teorica, IFT-UAM/CSIC,
Cantoblanco, 28049, Madrid, Spain}

\author{J. M. No}
\email[]{josemiguel.no@uam.es}
\affiliation{Departamento de Fisica Teorica, Universidad Autonoma de Madrid,
Cantoblanco, 28049, Madrid, Spain
}
\affiliation{Instituto de Fisica Teorica, IFT-UAM/CSIC,
Cantoblanco, 28049, Madrid, Spain}

\author{V. Sanz}
\email[]{v.sanz@sussex.ac.uk}
\affiliation{
Department of Physics and Astronomy, University of Sussex, BN1 9QH Brighton, United Kingdom}

\author{J. F. de Troc\'oniz}
\email[]{jorge.troconiz@uam.es}
\affiliation{Departamento de Fisica Teorica, Universidad Autonoma de Madrid,
Cantoblanco, 28049, Madrid, Spain
}

%%%%%%%%%%%%%%%%%%%%%%%%%%%%%%%%%%%%%%%%%%%%%%%%%%%%%%%%%%%%%%%
\hfill\draft{ }

\begin{abstract}
We propose a new collider probe for axion-like particles (ALPs), and more generally for pseudo-Goldstone bosons: 
non-resonant searches which take advantage of the derivative nature of their interactions with Standard Model particles.
ALPs can participate as off-shell mediators in the $s$-channel of $2 \to 2$ scattering processes at colliders like the LHC. 
We exemplify the power of this novel type of search 
by deriving new limits on ALP couplings to gauge bosons via the processes $p p \to Z Z$, $p p \to \gamma \gamma$  and $p p \to j j$ using Run 2 
CMS public data, probing previously unexplored areas of the ALP parameter space. 
In addition, we propose future non-resonant searches involving the ALP coupling to other electroweak bosons and/or the Higgs particle.

\end{abstract}

\maketitle

%%%%%%%%%%%%%%%%%%%%%%%%%%%%%%%%%%%%%%%%%%%%%%%%%%%%%%%%%%%%%%%%%%%%%%%%%%%%%%%%%%%%%%%%%%
%                                    Content                                             % 
%%%%%%%%%%%%%%%%%%%%%%%%%%%%%%%%%%%%%%%%%%%%%%%%%%%%%%%%%%%%%%%%%%%%%%%%%%%%%%%%%%%%%%%%%%

%%%%%%%%%%%%%%%%%%%%%%%%%%%%%%%%%%%%%%%%%%%%%%%%%%%%%%%%%%%%%%%%%%%%%%%%%%%%%%%%%%%%%%%%%%
\section{I. Introduction} 
%%%%%%%%%%%%%%%%%%%%%%%%%%%%%%%%%%%%%%%%%%%%%%%%%%%%%%%%%%%%%%%%%%%%%%%%%%%%%%%%%%%%%%%%%%
%\linenumbers

\vspace{-3mm}

Axion-like particles (ALPs)~\cite{Georgi:1986df,Choi:1986zw}, and more generally pseudo-Goldstone bosons,  
often appear in extensions of the Standard Model (SM). They may be connected to solutions to the 
strong CP problem~\cite{Peccei:1977hh,Peccei:1977ur,Wilczek:1977pj,Weinberg:1977ma,Rubakov:1997vp,Berezhiani:2000gh,Hsu:2004mf,Hook:2014cda,Fukuda:2015ana,
Chiang:2016eav,Dimopoulos:2016lvn,Gherghetta:2016fhp,Kobakhidze:2016rwh,Agrawal:2017ksf,Agrawal:2017evu,Gaillard:2018xgk,Gavela:2018paw}  
and/or to the existence of new spontaneously-broken global symmetries in Nature. In the following the term ALP will be used indistinctly 
to denote all such pseudo-scalars. 

ALPs are being searched at high-energy
colliders~\cite{Jaeckel:2012yz,Mimasu:2014nea,Jaeckel:2015jla,Knapen:2016moh,Brivio:2017ije,Bauer:2017ris,Mariotti:2017vtv,Baldenegro:2018hng,Craig:2018kne,Ebadi:2019gij},  
beam dump experiments~\cite{Dobrich:2015jyk,Dobrich:2019dxc}, via their effects in flavour 
physics~\cite{Izaguirre:2016dfi,Dobrich:2018jyi,CidVidal:2018blh,Aloni:2018vki,Gavela:2019wzg,Merlo:2019anv} and through their 
astrophysical signatures~\cite{Anastassopoulos:2017ftl,Armengaud:2014gea,Payez:2014xsa,Jaeckel:2017tud} (see Ref.~\cite{Essig:2013lka} for a review).

\begin{figure}[h]
\begin{center}
\includegraphics[width=0.185\textwidth]{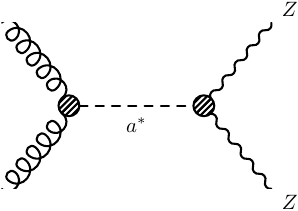}
\hspace{6mm}
\includegraphics[width=0.185\textwidth]{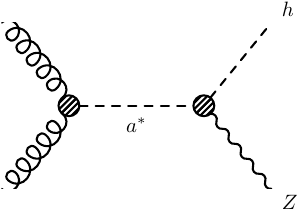}
\vspace{-3mm}
\caption{\small Feynman diagrams for the processes $g g \to Z Z$ (left) and $g g \to Z h$ (right) via an off-shell ALP in the $s$-channel.
}
\label{Fig_Feynman}
\end{center}
\end{figure}

In this work we propose a novel approach to probe the existence of ALPs at high-energy colliders, namely non-resonant searches where the 
ALP is an off-shell mediator in the $s$-channel of $2 \to 2$ scattering processes. The ALP pseudo-Goldstone nature implies that its interactions 
with SM particles are dominantly derivative, enhancing the cross sections for center-of-mass energies $\hat{s} \gg m_a^2$, where $m_a$ denotes the mass of the ALP $a$.   
In this kinematical regime, the processes tailored to search for ALPs at the Large Hadron Collider (LHC)
include those with two SM bosons in the final state:~electroweak 
gauge bosons ($W$, $Z$, $\gamma$), gluons $g$ and/or the Higgs particle $h$.
For $m_a \ll$ 100 GeV, the gluon-initiated $2 \to 2$ di-boson scattering processes $p p\,(gg) \to Z Z,$ $W W$, $Z \gamma$ and $Z h$ may be mediated by 
a virtual ALP, as shown in Fig.~\ref{Fig_Feynman}. This can also occur for the processes
$ p p\,(gg) \to j j\, (g g)$ or $ p p\,(gg) \to \gamma \gamma$ in the regime where a large invariant mass $m_{jj}$ or $m_{\gamma\gamma}$ 
is required in the final state.

The theoretical framework used throughout this work is the model-independent approach of effective field theories (EFT). 
If the Higgs particle is considered to be part of an exact $SU(2)_L$ doublet at low energies, as predicted in the SM, the putative beyond the Standard Model (BSM) 
electroweak physics may then be described by an EFT {\it linear} expansion~\cite{Buchmuller:1985jz,Grzadkowski:2010es} 
in terms of towers of gauge invariant operators  
ordered by their mass dimension. 
Alternatively, since a non-doublet component of the Higgs particle is at present experimentally allowed (at the $\sim 10\%$ level~\cite{ATLAS:2019slw}),
a {\it non-linear} EFT (also called {\it chiral})~\cite{Feruglio:1992wf,Alonso:2012jc,Azatov:2012bz,Alonso:2012px,Alonso:2012pz,Buchalla:2013rka,Brivio:2013pma} 
based on a momentum expansion is also of interest.  
In the following we concentrate on the linear EFT for the SM plus an ALP~\cite{Georgi:1986df,Choi:1986zw, Brivio:2017ije}, 
and discuss when pertinent the comparison with a chiral EFT, notably for the interactions between the ALP and the Higgs boson~\cite{Brivio:2017ije,Bauer:2017ris}.

\vspace{-12mm}

\section{II. Bosonic ALP Lagrangian}
\label{sec:EFT}

\vspace{-3mm}

\noindent {\bf\textit{Linear expansion.}} 
In the linear ALP EFT, 
the new physics scale to be considered is the ALP decay constant $f_a$, which will 
weight down the higher-dimensional operators built from the SM fields and $a$. 
The most general CP-conserving effective Lagrangian describing  {\it bosonic} ALP couplings contains -- up to next-to-leading 
order (NLO) -- only four independent operators of mass dimension five~\cite{Georgi:1986df,Choi:1986zw,Salvio:2013iaa,Brivio:2017ije},  
%%%%%%%%%%%%%
\begin{align}
\begin{split}
\delta \mathcal{L}_{\mathrm{eff}} &\supset 
c_{\tilde{G}}\, \mathcal{O}_{\tilde{G}}+c_{\tilde{B}}\, \mathcal{O}_{\tilde{B}}+c_{\tilde{W}} \,\mathcal{O}_{\tilde{W}}\, +{c_{a\Phi}} \,\mathcal{O}_{a\Phi}\,,
\end{split}
\label{eq:lagrangian-0}
\end{align}
%%%%%%%%%%%%%
\noindent where 
%%%%%%%%%%%%%
\begin{align}
\begin{split}
\mathcal{O}_{\tilde{G}} \equiv - \dfrac{a}{f_a}G_{\mu\nu} \widetilde{G}^{\mu\nu}\,, \quad &  
\mathcal{O}_{\tilde{W}} \equiv - \dfrac{a}{f_a}W_{\mu\nu}^a \widetilde{W}^{\mu\nu}_a \,,\\[0.4em]
\mathcal{O}_{\tilde{B}} \equiv - \dfrac{a}{f_a}B_{\mu\nu} \widetilde{B}^{\mu\nu} \,, \quad &
\mathcal{O}_{a\Phi}  \equiv i\dfrac{\partial^\mu a}{f_a}\, \Phi^\dagger \overleftrightarrow{D}_\mu \Phi 
\,.
\label{operators}
\end{split}
\end{align}
%%%%%%%%%%%%%
%
\noindent 
In these expressions  
$G_{\mu\nu}$, $W_{\mu\nu}$ and $B_{\mu\nu}$  denote respectively the $SU(3)_c\times SU(2) \times U(1)$  field strengths, 
and the dual field strengths are defined as $\tilde{X}^{\mu\nu}\equiv \frac12 \epsilon^{\mu\nu\rho\sigma}X_{\rho\sigma}$, with $\varepsilon^{0123}=1$.  
The $c_i$ constants are real operator coefficients and $\Phi$ denotes the SM Higgs doublet,
with $\Phi \overleftrightarrow{D}_\mu \Phi \equiv \Phi^\dagger \big{(}D_\mu\Phi\big)-\big{(}D_\mu \Phi\big{)}^\dagger \Phi {}$.
The first three operators in Eq.~(\ref{eq:lagrangian-0}) induce physical $a g g$, $a\gamma \gamma$, $a\gamma Z$, $a Z Z$ and $a W^+ W^-$ interactions, 
\begin{eqnarray}
\delta \mathcal{L}_{\mathrm{eff}} &\supset& - \frac{g_{agg}}{4}a\,G_{\mu\nu} \widetilde{G}^{\mu\nu}  -
\frac{g_{a\gamma\gamma}}{4}a\,F_{\mu\nu} \widetilde{F}^{\mu\nu} - \frac{g_{aZ\gamma}}{4}a\,F_{\mu\nu} \widetilde{Z}^{\mu\nu} \nonumber \\
&-& \frac{g_{aZZ}}{4}a\,Z_{\mu\nu} \widetilde{Z}^{\mu\nu} - \frac{g_{aWW}}{4}a\,W_{\mu\nu} \widetilde{W}^{\mu\nu}\,,
\label{eq:lagrangian_gs}
\end{eqnarray}
 where 
 \begin{align}
\label{gagluon}
g_{agg} & =\frac{4}{ f_{a}}\,c_{\tilde{G}} \,, \quad
g_{a\gamma\gamma} = \frac{4}{f_{a}} \, \big(s_w^2 \, c_{\tilde{W}}+ c_w^2 \, c_{\tilde{B}} \big)\\
g_{aWW} &= \frac{4}{ f_a}  \, c_{\tilde{W}} \,,\quad 
g_{aZZ} = \frac{4}{ f_a} \,({c_w^2}\, c_{\tilde{W}}+ {s_w^2}\, c_{\tilde{B}}) \label{gaZZ}\\
\label{gaZgamma}
g_{a\gamma Z} &= \frac{8}{ f_a}  \,s_wc_w ( c_{\tilde{W}}-  c_{\tilde{B}} )\,,
\end{align}
and  $s_w$ and $c_w$ denote respectively the sine and cosine of the Weinberg mixing angle. 
The Feynman rule for the interaction $a V_1 V_2$ (with $V_{1,2}$ being SM gauge bosons) stemming from these operators is given by 
\begin{equation}
\label{Feynman_Rule}
- i \,g_{a V_1 V_2}\, p^{\rho}_{V_1} p^{\sigma}_{V_2} \epsilon_{\mu\nu\rho\sigma}  \,.  
\end{equation}
The last operator in Eq.~(\ref{operators}), $\mathcal{O}_{a\Phi}$, induces a mixing between $a$ and the would-be Goldstone boson eaten by the $Z$.
Its physical impact is best illustrated via a Higgs field redefinition,  
$\Phi \to \Phi \, e^{i c_{a\Phi} a/f_a}$~\cite{Georgi:1986df}, which  trades $\mathcal{O}_{a\Phi}$ for %%%%%%%%%%%%%%%%
\begin{align}
\label{eq:leff-alp}
i \dfrac{a}{f_a} \Big{[}\overline{Q} Y_u \widetilde{\Phi} u_R -  
\overline{Q} Y_d \Phi d_R - \overline{L} Y_\ell \Phi \ell_R\Big{]}+\mathrm{h.c.}\,,
\end{align}
%%%%%%%%%%%%%%%% 
where $Y_{u,d,\ell}$ denote the SM Yukawa matrices.
We focus in this Letter on experimental signals involving ALPs and SM bosons ($W,\,Z,\,\gamma,\,g$ and $h$), yet we briefly comment 
on signatures involving the $\mathcal{O}_{a\Phi}$ fermionic coupling in Sect. IV-4.\footnote{A complete --bosonic and fermionic-- ALP basis can be obtained 
substituting $\mathcal{O}_{a\Phi}$ in Eq.~(\ref{operators}) by general flavour-changing 
fermionic operators e.g.~\cite{Georgi:1986df,Choi:1986zw, Salvio:2013iaa, Brivio:2017ije}. The physical effects of the latter are proportional to 
the Yukawa couplings of the fermions involved.}

\vspace{3mm}

\noindent {\bf\textit{Chiral expansion.}} 
The operators $\mathcal{O}_{\tilde{G}}$, $\mathcal{O}_{\tilde{W}}$ and $\mathcal{O}_{\tilde{B}}$ in Eq.~(\ref{operators}) also 
appear in the chiral expansion at NLO.
In addition, and at variance with the linear EFT, novel ALP-Higgs couplings are present  
in the chiral expansion already at leading order (LO), namely the operator $\A_{2D}(h)$~\cite{Brivio:2017ije} which is 
a custodial breaking two-derivative operator with mass dimension three: 
\beq
\LL^{\text{LO}}_a \supset 
c_{2D}\A_{2D}(h) = c_{2D} \left[iv^2\tr[\T\V_\mu]\de^\mu\frac{a}{f_a}\F(h) \right]\,,
\label{La_LO}
\eeq
where  
  $v = 246$ GeV denotes the electroweak scale as defined from the $W$ mass, 
$\V_\mu(x)\equiv \left(\DLR_\mu\U(x)\right)\U(x)^\dag$,  $\T(x)\equiv \U(x)\sigma_3\U(x)^\dag$ and $\U(x)=e^{i\sigma_j \pi^j(x)/v}\,$, 
with $\pi^j(x)$ corresponding to the longitudinal degrees of freedom of the electroweak gauge bosons and $\sigma_j$ are the Pauli matrices. 
The physical Higgs particle $h$ is introduced in the chiral expansion via 
polynomial functions~\cite{Grinstein:2007iv} of $h/v$, 
$\cF(h)=1+ a_{2D} h/v + b_{2D}(h/v)^2+ \mathcal{O}(h^3/v^3)$,
with $a_{2D}$, $b_{2D}$ constant coefficients. 
$\A_{2D}$ is the chiral counterpart (``sibling") of the linear operator $\mathcal{O}_{a\Phi}$ in Eq.~(\ref{operators}), 
with a key difference: in addition to ALP-fermion couplings analogous to those in Eq.~\eqref{eq:leff-alp}, 
$\A_{2D}$ induces interactions between the ALP, the electroweak gauge bosons and any number of Higgs particles, e.g.~a 
trilinear $a-Z-h$ coupling (see Fig.~\ref{Fig_Feynman}). 
The associated experimental signatures at the LHC will be discussed in Sect.IV-4.
Note that such couplings can be found in the linear expansion only at next-to-NLO (NNLO ), i.e. mass dimension seven~\cite{Bauer:2017ris,Bauer:2016zfj},
and are thus expected to yield subleading effects in that case.

\vspace{-0.5cm}

\section{III. Non-resonant ALP-Mediated Di-boson production}

\vspace{-3mm}

The key observation is that, due to the derivative nature of the ALP interactions under discussion, 
the ALP mediated scattering processes $g g \to a^* \to V_1 V_2$ exhibit a harder scaling with the invariant mass of the event $\sqrt{\hat{s}} = m_{V_1 V_2}$ than 
usual $s$-channel  mediated exchanges.

 In all generality, the contributions from bosonic ALP couplings in Eq.~\eqref{eq:lagrangian_gs} 
 interfere with the absortive part of   SM  $2 \to 2$ di-boson amplitudes.   Nevertheless, given the present loose bounds on ALP couplings, pure ALP exchange dominates the cross section for some LHC channels. A quartic dependence on ALP couplings results in this case, 
\begin{equation}
\label{ALP_XS}
\sigma_{V_1 V_2} \propto g^2_{agg}\, g^2_{a V_1 V_2}\, {\hat s}\,\sim \frac{\hat{s}}{f_a^4},
\end{equation}
in the ALP off-shell regime $\hat{s} \gg m_a^2, m_{V_i}^2$.\footnote{This has been noted in a different setup in Ref.~\cite{Baldenegro:2018hng}.} 
The same type of energy behaviour holds for $g g \to a^* \to Z h$ from Eq.~(\ref{La_LO}). 
Such energy dependence is valid only as far as the energies probed in the scattering process are smaller than the cut-off 
scale of the EFT, $\sqrt{\hat{s}} < f_a$. 

The behaviour in Eq.~\eqref{ALP_XS} 
is to be compared with the energy dependence for a usual $2\to 2$ $s$-channel mediated process, which scales instead as $1/\hat{s}$ far above 
from the $s$-channel resonance.
Factoring in the proton parton distribution functions (PDFs), which tame the energy growth in Eq.~\eqref{ALP_XS}, the differential cross section for 
the ALP-mediated process $p p \to a^{*} \to V_1 V_2$ diminishes --at energies much larger 
than the resonance's mass-- more slowly with the invariant mass  
than for a usual $s$-channel resonance whose couplings do not depend on the momenta involved. The 
momentum dependence of the ALP interaction in Eq.~\eqref{Feynman_Rule} significantly smooths out 
the decrease of the cross section at large $\sqrt{\hat{s}}$, allowing to distinguish ALP-mediated processes from the SM background.

For sufficiently small  ALP couplings, the size of the interference with the SM background 
becomes comparable to the pure ALP-signal and must be taken into account. 
The value of the couplings for which this happens depends on the specific final state $V_1 V_2$. For 
$\gamma\gamma$ and the other electroweak di-boson final states,  the ALP signal interferes with one-loop SM processes. 
 The interference is constructive or destructive depending on the relative sign of $g_{agg}$ and $g_{a V_1 V_2}$;  
in any case, it can be disregarded  at present, given the size of ALP couplings that can be currently probed at the LHC (see section~IV).
 Future LHC analyses will need to include them, though, and this exploration is deferred to a future work~\cite{GNST}.

The situation is different for di-jet ($gg$) final states: the SM contributes at tree-level (via gluon exchange), and the interference already dominates at present over the pure ALP signal, for the $g_{agg}$ values at reach. It is destructive and will be taken into account in the  $g g  \to gg$ analysis in Sect.~IV-3.

\begin{figure}[h]
\begin{center}
\includegraphics[width=0.46\textwidth]{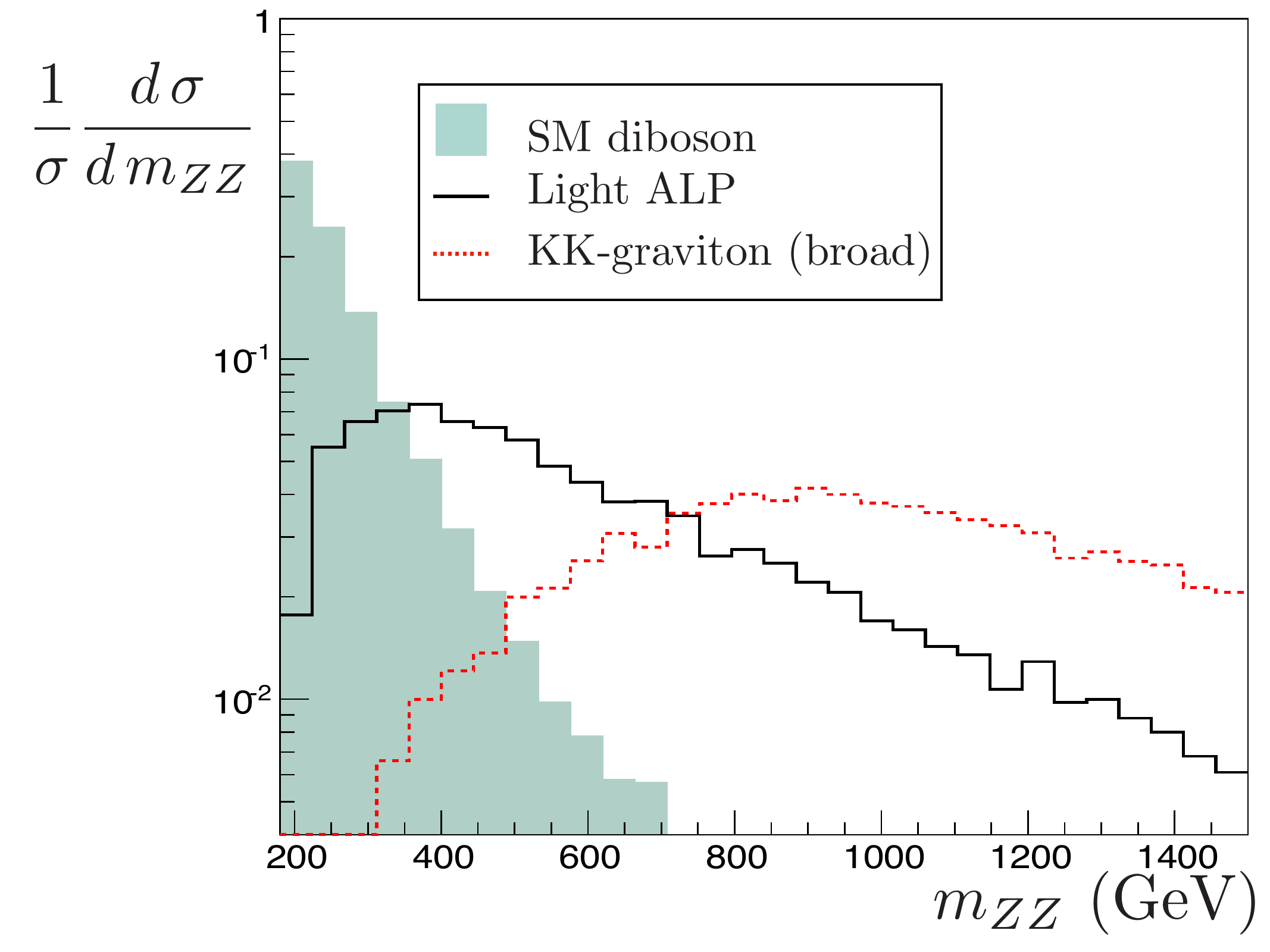}

\vspace{-2mm}

\caption{\small 13 TeV LHC normalized differential cross section for di-boson production as a function of the di-boson invariant mass $m_{ZZ} = \sqrt{\hat{s}}$. 
The signal of an $s$-channel ALP with $m_a \ll m_{jj}$ (solid black line) is compared to the SM prediction (green area) and a broad KK-graviton with mass 500 GeV and 30\% width  (dashed red line). 
}
\label{Fig_dijet}
\end{center}
\end{figure}

It is   illustrative to compare the shape of the $s$-channel invariant mass distribution stemming from an off-shell  ALP of negligible mass,  
with that from a heavier  (and broad) mediator with derivative couplings, e.g. a KK-graviton with large width.
 This is illustrated in Fig.~\ref{Fig_dijet} for an electroweak  diboson (ZZ)  final state; the SM expectation is depicted in addition.  
The different generic patterns of decelerated decrease highlight  that it is possible to 
distinguish  the presence of a light ALP from that of a heavy BSM resonance, even if the heavy resonance's couplings are derivative as well,  via the study of $m_{V_1 V_2}$  distributions.

A supplementary handle to discern whether a putative BSM signal would correspond to an ALP is given 
by the angular distribution of the final states~\cite{Gao:2010qx}: the rich Lorentz structure of ALP couplings in Eq.~\eqref{Feynman_Rule} induces a 
distinctive angular pattern\footnote{See e.g. Refs.~\cite{Hagiwara:2009wt,Berge:2008wi,Dolan:2014upa,Boudjema:2015nda} for analogous studies in Higgs physics.} which may 
be used to infer the pseudoscalar nature of $a$. We leave such an exploration for a future work. 

\vspace{2mm}

The non-resonant $s$-channel ALP signatures explored in this work have several further attractive features: 
{\it(i)}~In the regime under discussion with $\hat{s} \gg m_a^2$, the signal cross section and distributions are essentially independent of the value of 
$m_a$. This implies that the search is equally sensitive to any $m_a$ significantly below the energy range probed by the search. In particular,
for the LHC searches considered in Sect.~IV, the derived sensitivity can be safely applied to any ALP mass below $100$ GeV.
{\it(ii)}~Being a non-resonant process, no hypothesis is needed on the value of other possible couplings which do not contribute to the process 
under consideration. This is at variance with on-shell analyses, for which the dependence on other  ALP couplings may appear through the partial  
decay widths.\footnote{Most present ALP limits based on resonant processes have considered only one independent $g_{a V_1 V_2}$  coupling at a time among 
the set in Eq.~\eqref{eq:lagrangian_gs} (see e.g. Refs.~\cite{Mimasu:2014nea,Jaeckel:2015jla,Knapen:2016moh,Dobrich:2015jyk,Izaguirre:2016dfi}), with some recent analyses 
considering the simultaneous presence of  two or at most three independent couplings~\cite{Mariotti:2017vtv,Alonso-Alvarez:2018irt,Gavela:2019wzg}.} 
In this sense, non-resonant searches are more model-independent and thus more robust. 

From the theoretical point of view, the  $g_{a V_1 V_2}$ couplings   depend only on the ratio $c_i/f_a$ (see Eqs.~\eqref{gagluon}-\eqref{gaZgamma}), 
but the value of $f_a$ is relevant to assess the validity of the EFT, which  
limits the energy range  that can be safely considered in an 
LHC search (e.g. to bins satisfying $\sqrt{\hat{s}} < f_a$)\footnote{If the underlying BSM theory were in the weak coupling regime, and led at one-loop 
to the operators in Eq.~(\ref{operators}), their coefficients could plausibly be suppressed by an 
additional $ \alpha_i /(8\pi)$ factor. This would drastically reduce the set of valid energy bins in LHC searches. We stick here instead to the general 
and widespread definitions in Eqs.~(\ref{operators}) and~\eqref{eq:lagrangian_gs}.},   
and discuss this for specific LHC searches in Sect.~IV. 
Another pertinent question is the possible impact of  radiative corrections and of higher dimensional operators. 
For the former, self-energy corrections to the $s$-channel ALP propagator only become non-negligible close to the EFT validity boundary, and we do not consider 
their effect here.
Higher dimensional operators, e.g. those weighted down by the same $\mathcal{O} (1/f_a^2)$ factor than the amplitudes discussed above, can also 
contribute only at loop level,  as $f_a$ must intervene as powers of $a/f_a$ and no ALP is present in the final states considered here. 
Furthermore, only by engineered fine-tuned cancellations could such operators impact significantly on 
the results of this work. 
 
\vspace{-5mm} 
 
\section{IV. Non-resonant LHC searches}

\vspace{-3mm}

In this section we derive new limits on  $g_{a V_1 V_2}$ couplings through the non-resonant ALP-mediated processes discussed above, using public data from LHC Run 2 
 ($\sqrt{s} =$ 13 TeV) CMS searches.  ALP production in the $s$ channel is dominated by gluon-gluon fusion, as the $q \bar{q}$ induced ALP production amplitude is  proportional to the quark masses -- see Footnote 1-- and thus highly suppressed.  Possible final states to be considered  include $gg$, $ZZ$,  $WW$, $Z\gamma$, $\gamma \gamma$ or $Zh$. 
While it is of high interest to explore all of them, since they probe different operator combinations within the EFT, 
we focus below on the processes $p p \to a^* \to Z Z$, $p p \to a^* \to  \gamma \gamma$ and $p p \to a^* \to gg$. For these channels, the CMS collaboration
has recently published new results, providing explicit calculations of the corresponding SM backgrounds.   
We  use those public data to compute approximate limits on $g_{agg}\times g_{aZZ}$, $g_{agg}\times g_{a\gamma\gamma}$ and $g_{agg}$, respectively.
In all three analyses, the ALP mass is fixed to $m_a = 1$ MeV (i.e. effectively massless at LHC energies) and 
the ALP width $\Gamma_a$ is assumed to respect $\Gamma_a \ll m_a$.

For the $p p \to a^* \to Z Z$ and $p p \to a^* \to \gamma \gamma$ channels, our sensitivities are estimated from a simplified binned 
likelihood ratio analysis. The likelihood function is
built as a product of bin Poisson probabilities 
\begin{eqnarray}
L(\mu) = \prod_k \,e^{-(\mu s_k +\, b_k)}\, \frac{(\mu s_k + b_k)^{n_k}}{n_k !} \,,   
\label{likelihood_NS}
\end{eqnarray}
where $n_k$, $b_k$ and $s_k$ denote respectively the observed data, SM background and ALP signal prediction in a given  
bin $k$, and the signal strength modifier $\mu$ is taken as the only floating parameter in the likelihood fit (see Ref.~\cite{Brivio:2017ije} for details), 
with no systematic uncertainties considered for simplicity. For the $p p \to a^* \to g g$ channel we perform a $\chi^2$ fit to the data 
including systematic errors but no bin-to-bin correlations.

\vspace{1mm}

Other important search channels are also briefly discussed below, albeit their analysis is left for the future: 
$p p \to a^* \to  Z\gamma$, $p p \to a^* \to Z h$ (which provides a unique window into the chiral EFT via the  
operator~$\A_{2D}$ in Eq.~\eqref{La_LO}) and $p p \to a^* \to t \bar{t}$ (which would yield access to the 
operator $\mathcal{O}_{a\Phi}$ in Eq.~\eqref{operators}).  

\vspace{-5mm}

\subsection{1) $\mathbf{p p \to a^* \to Z Z}$} 
\label{sec:CMS_ZZ}
%%%%%%%%%%%%%%%%%%%%%%%%%%%%%%%%%%%%%%%%%%%%%%%%%%%%%%%%%%%%%%%%%%%%%%%%%%%%%%%%%%%%%%%%%% 

\vspace{-3mm}

The process $p p \to a^{*} \to Z Z \to \ell\ell q \bar{q}$ is studied next, following the semi-leptonic di-boson CMS analysis at LHC $\sqrt{s} = 13$ TeV with 
$35.9$ fb$^{-1}$~\cite{Sirunyan:2018hsl}.~We focus on the ``low-mass merged" CMS analysis 
category targeting the invariant mass region $m_{ZZ} \in [450, \,2000]$ GeV, with one $Z$ boson decaying leptonically, $Z \to \ell \ell$ and the other 
decaying hadronically. The boosted hadronic $Z$ decay products are required to merge into a single jet, $Z \to J$. The jet is 
reconstructed via the anti-$k_T$ algorithm with $R = 0.8$ (AK8).
Our signal process is simulated in {\tt MadGraph$\_$aMC@NLO}~\cite{Alwall:2014hca}, with a subsequent parton showering and hadronization 
with {\tt Pythia 8}~\cite{Sjostrand:2014zea} and detector simulation via {\tt Delphes 3}~\cite{deFavereau:2013fsa}, including 
the use of jet-substructure variables as discussed in detail in Appendix A1. 
Following Ref.~\cite{Sirunyan:2018hsl}, the analysis is divided into {\sl $b$-tagged} and {\sl untagged} categories, targeting respectively 
the $Z \to b \bar{b}$ and $Z \to q \bar{q}$ (with $q = u,d,s,c$) decays. The {\sl $b$-tagging} of the merged jet $J$ 
provides a strong background suppression, yielding a further increase in sensitivity.

\begin{figure}[h]
\begin{center}

\includegraphics[width=0.48\textwidth]{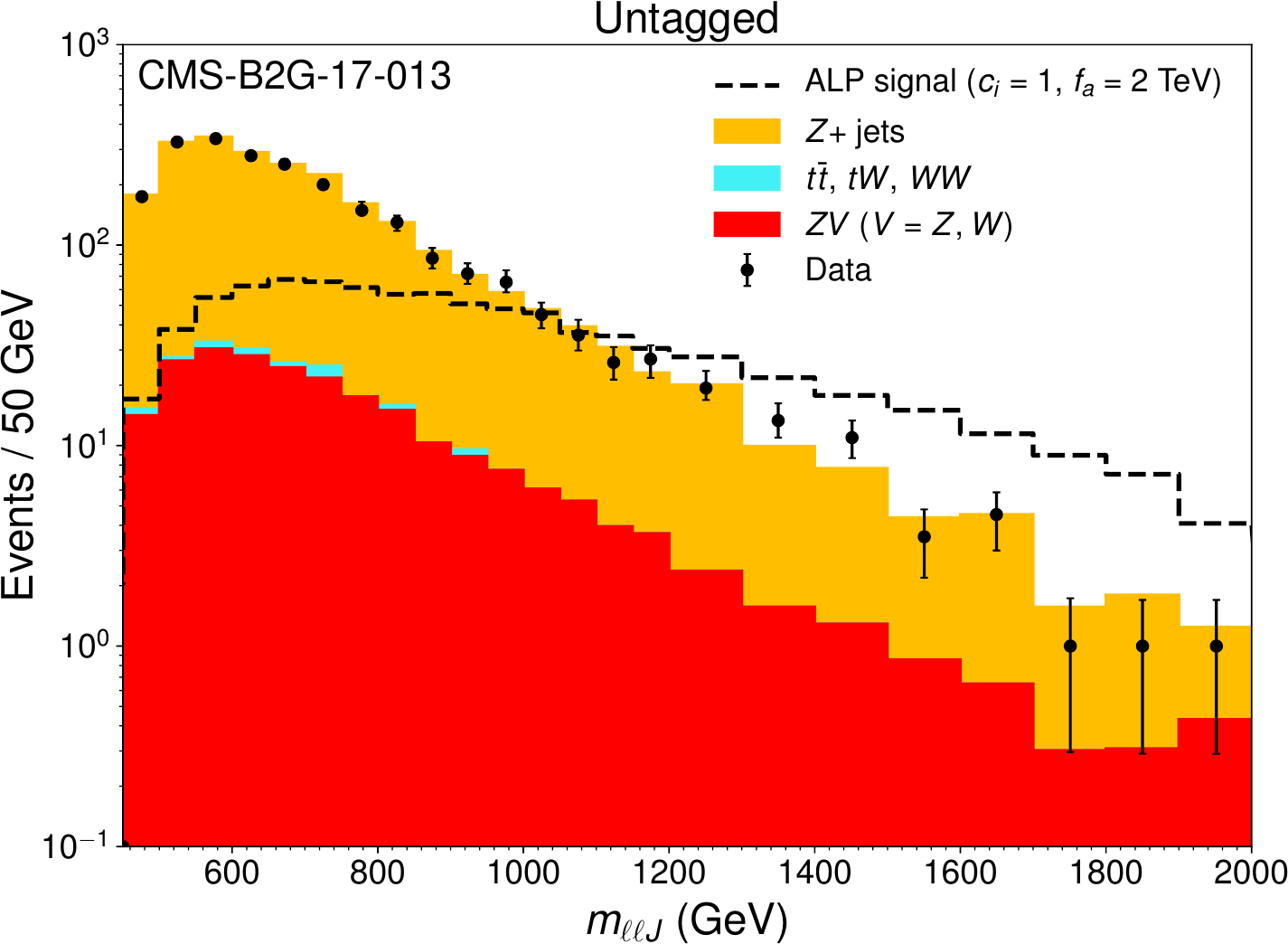}

\vspace{1mm}

\includegraphics[width=0.48\textwidth]{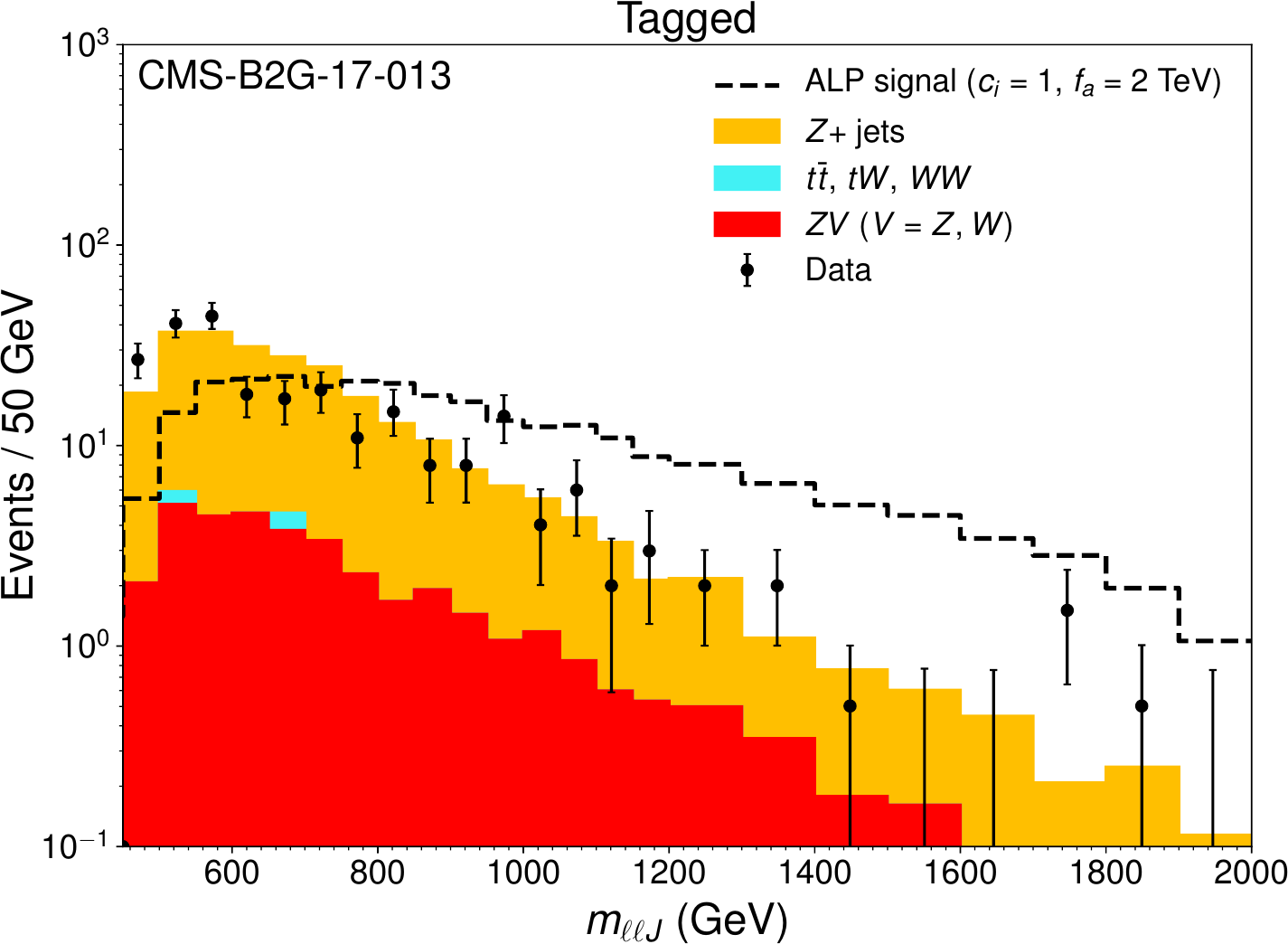}

\vspace{-2mm}

\caption{\small $m_{\ell \ell J}$ distributions for the ALP $ZZ$ signal with 
$c_i = 1$, $f_a = 2$ TeV (dashed black line) and SM background from $Z +$ jets (yellow), $ZV$ (red) and $t \bar{t}$ (cyan)
 after CMS event selection, in the {\sl untagged} (top) and {\sl $b$-tagged} (bottom) categories. The experimental data are shown as black dots.}
\label{Fig_DistSignalmZZ}
\end{center}

\vspace{-4mm}

\end{figure}

\begin{figure}[t]
\begin{center}
\includegraphics[width=0.48\textwidth]{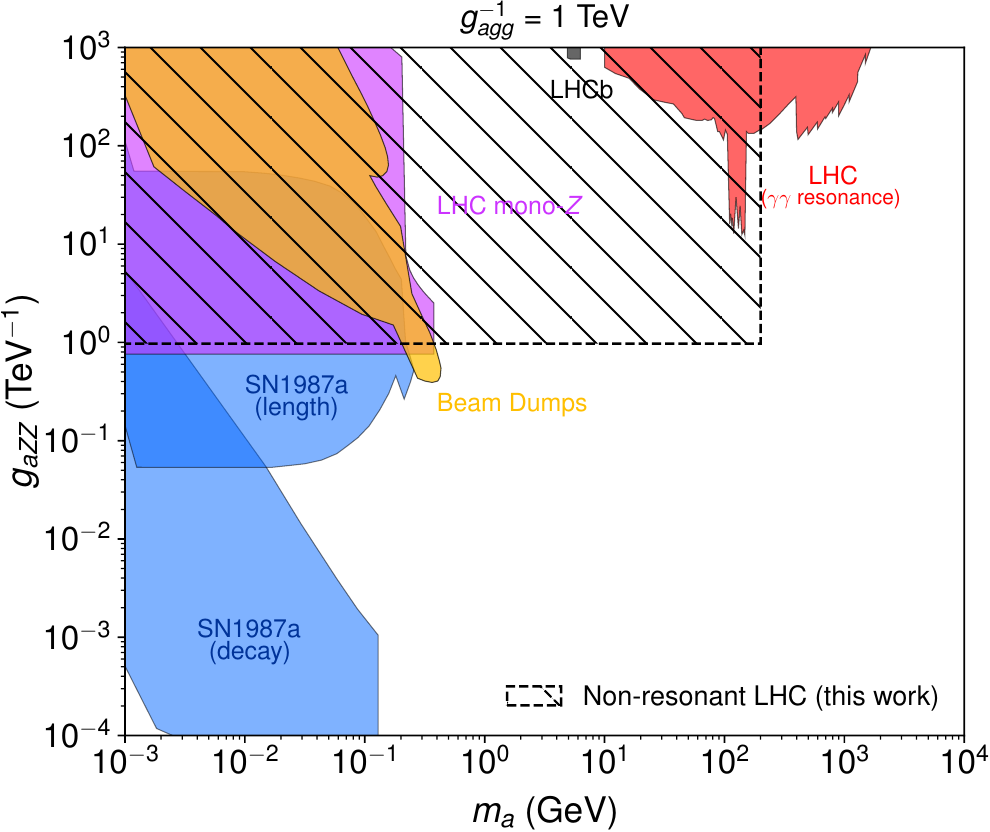}
\vspace{3mm}

\includegraphics[width=0.48\textwidth]{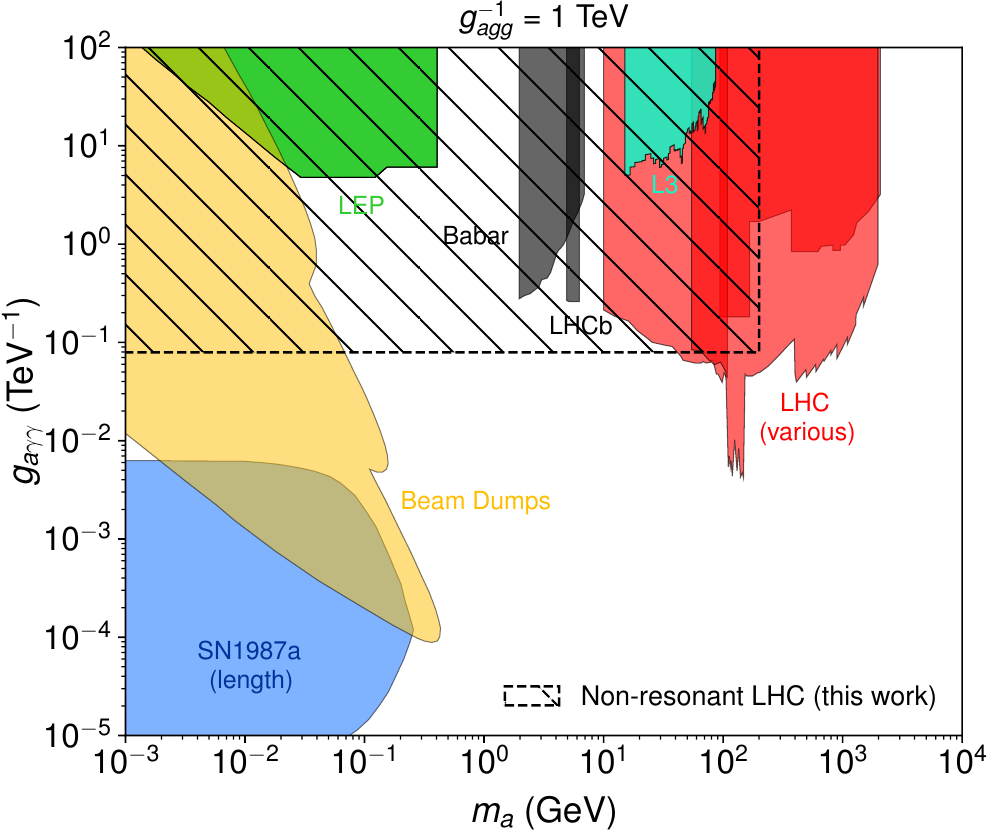}
\vspace{-2mm}

\caption{\small Top: Bounds on the ALP coupling $g_{aZZ}$ as a function of $m_a$. The hatched region corresponds to the limit from non-resonant 
LHC searches derived in this work using CMS di-boson data~\cite{Sirunyan:2018hsl}. Also shown are limits from LHC mono-$Z$ searches (violet), 
beam-dump experiments (yellow), supernova SN1987a (blue), LHCb (dark grey) and LHC resonant $\gamma\gamma$ searches (red), see text for details.
Bottom: Bounds on the photonic couplings $g_{a\gamma\gamma}$, with color code as for the top figure. Limits from BaBar (dark grey), L3 (cyan) and 
LEP (green) are also depicted, see text for details.}
\label{Fig_Bounds}
\end{center}

\vspace{-4mm}

\end{figure}

As an illustration of the impact of the  derivative nature of ALP interactions, the $\sqrt{s} = 13$ TeV cross-section $\sigma(p p \to a^* \to Z Z)$ 
for $c_{\tilde G} = c_{\tilde W} = c_{\tilde B} = 1$ and $f_a = 1$ TeV is $81$ pb. 
The CMS event selection is discussed in detail in Appendix A1.
Fig.~\ref{Fig_DistSignalmZZ} shows the invariant mass $m_{\ell\ell J}$ distribution resulting for the signal after the CMS event selection, for 
$c_i = 1$ and $f_a = 2$ TeV (corresponding to the largest value of $m_{\ell\ell J}$ in the CMS analysis), 
together with the SM background publicly available in Ref.~\cite{Sirunyan:2018hsl} (and dominated by  
$Z +$ jets), both for the {\sl untagged}  (top plot) and {\sl $b$-tagged} (bottom plot) categories. 
A binned likelihood analysis of the $m_{\ell\ell J}$ distribution after CMS event selection combining the 
{\sl untagged} and {\sl $b$-tagged} categories is then performed, which allows to set a 95\% C.L. exclusion limit on the signal cross section 
of $\sigma = 25$ fb. This corresponds to $f_a > 4.1$ TeV for $c_i = 1$, and is valid for any value of the ALP mass up  $m_a \sim 200$ GeV 
without significant modifications of the signal properties. Note that, since the ``low-mass merged" CMS analysis uses data up to $m_{ZZ} = 2$ TeV, our  
derived limit on $f_a$ for  $c_i = 1$ lies within the region of validity of the EFT. In Fig.~\ref{Fig_Bounds} (top) the corresponding new 
limit on $g_{aZZ}$ (see Eq.~\eqref{gaZZ}) resulting from our non-resonant analysis is depicted as a hatched area, for a fixed value $g_{agg}^{-1} = 1$ TeV.

For comparison, Fig.~\ref{Fig_Bounds} (top) depicts as well previous bounds in the literature for $g_{aZZ}$,  which also assume the additional presence of $g_{agg}$, albeit obtained from on-shell ALP searches.  For $m_a \lesssim 0.1$ GeV, the ALP is stable on LHC scales, resulting in constraints on $g_{aZZ}$ from mono-$Z$ searches (in violet), see Ref.~\cite{Brivio:2017ije}. The  radiative (2-loop) contribution of $g_{aZZ}$ to $g_{a\gamma\gamma}$ allows to obtain further constraints  for certain ranges of ALP masses for which strong constraints on $g_{a\gamma\gamma}$ exist (see the discussion in Refs.~\cite{Bauer:2017ris, Alonso-Alvarez:2018irt}). For ALP masses below the GeV scale,  limits on $g_{aZZ}$ are thus set by  beam dump searches (in yellow)~\cite{Riordan:1987aw,Bjorken:1988as,Blumlein:2013cua} (we adapt here the bounds compiled in Ref.~\cite{Dobrich:2015jyk}), 
and by energy-loss arguments applied to the supernova SN1987a~\cite{Payez:2014xsa,Jaeckel:2017tud} (in blue), both through absence of extra cooling (labelled ``length" 
in Fig.~\ref{Fig_Bounds} and through absence of a photon burst from decaying emitted axions (labelled ``decay" in Fig.~\ref{Fig_Bounds}. 
Furthermore, the radiative contribution of $g_{aZZ}$ to $g_{a\gamma\gamma}$   is also constrained by LHCb~\cite{Benson:2314368} (see Ref.~\cite{CidVidal:2018blh}) in the small region $4.9$ GeV $< m_a < 6.3$ GeV (in dark grey) and by ATLAS/CMS searches for $\gamma\gamma$ resonances (in red) for $m_a > 10$ GeV (we adapt here the bounds from Refs.~\cite{Jaeckel:2012yz,Mariotti:2017vtv}). We stress that the latter limits are from LHC Run 1 ($\sqrt{s} = 7$ and $8$ TeV), and as such $\sqrt{s} = 13$ TeV Run 2 analyses should significantly improve on those.  Next, although LHC tri-boson searches for $m_a \gg 100$ GeV have yielded  very weak constraints~\cite{Craig:2018kne},  the radiative contribution of $g_{aZZ}$ to $g_{a\gamma \gamma}$ provides  as well sizeable constraints.  We do not include here, though, the expected tree-level bounds on $g_{aZZ}$ from $ZZ$ resonance searches by ATLAS and CMS (e.g.~from Ref.~\cite{Sirunyan:2018hsl})  for $m_a > 200$ GeV. To our knowledge, these have not yet been obtained and are complementary to the non-resonant search presented in this work. The study of such $ZZ$ resonant searches is left for a forthcoming work~\cite{GNST}.

\vspace{-5mm}

\subsection{2) $\mathbf{p p \to a^* \to \gamma \gamma}$} 
\label{sec:CMS_gammagamma}

\vspace{-3mm}

Non-resonant ALP searches are also possible for final states to which a light ALP could decay, such as $\gamma \gamma$,  by 
selecting events with a large invariant mass $m_{\gamma\gamma} \gg m_a$. The recent CMS search for non-resonant new physics 
in $\gamma\gamma$ final states~\cite{Sirunyan:2018wnk} with 35.9 fb$^{-1}$ of $13$ TeV LHC data is used here.
In analogy with the previous section, we simulate the signal process $p p \to a^* \to \gamma \gamma$  with {\tt MadGraph$\_$aMC@NLO}, {\tt Pythia 8} and 
{\tt Delphes 3}, obtaining a signal cross-section $\sigma (p p \to a^* \to \gamma\gamma) = 47$ pb 
for $c_{\tilde G} = c_{\tilde W} = c_{\tilde B} = 1$, $f_a = 1$ TeV and with the initial requirement $m_{\gamma\gamma} > 500$ GeV.
The subsequent CMS event selection applied here is detailed in Appendix A2, with the 
main SM backgrounds~\cite{Sirunyan:2018wnk} 
being $\gamma\gamma$ and $\gamma + j$  
(with the jet $j$ mis-identified as a photon). 
After the event selection,  
we perform a binned likelihood analysis of the $m_{\gamma\gamma}$ distribution for the two selection categories discussed in  Appendix A2 according to the rapidity of the photons. This leads to a combined 95\% C.L. observed exclusion limit on the signal cross section of $\sigma \simeq 1.2$ fb. This limit corresponds 
to $f_a > 14.2$ TeV for $c_i = 1$, which we find to be valid up to $m_a \sim 200$ GeV without significant modifications of the signal properties. The resulting bound on $g_{a\gamma\gamma}$ is depicted in Fig.~\ref{Fig_Bounds} (bottom) for $g_{agg}^{-1} = 1$ TeV as a hatched area.  
Bounds from resonant searches at the LHC, beam dump experiments and astrophysical constraints (supernova SN1987a) are also shown, see Sect. IV-1 for details. 
For comparison, the figure also shows bounds from resonant searches by BaBar~\cite{Lees:2011wb} (in dark grey) (as obtained from Ref.~\cite{CidVidal:2018blh}), from 
L3~\cite{Adriani:1992zm} (in cyan), as well as  from LEP searches (in green) for new physics in $e^+ e^- \to 2\gamma, 3\gamma$ processes 
(see Refs.~\cite{Jaeckel:2015jla,Mimasu:2014nea} for a detailed discussion). Regarding the latter, Refs.~\cite{Jaeckel:2015jla,Mimasu:2014nea} assume 
a vanishing gluonic coupling $g_{agg}$, and thus apply in our case only for $m_a < 3\,m_{\pi}$; for $m_a > 3\,m_{\pi}$ the ALP can decay into hadronic final states
in the presence of a non-vanishing $g_{agg}$ coupling, which could significantly weaken the LEP bounds, and we refrain to claim an exclusion in that region. 
We also do not include here projected limits on $g_{a\gamma\gamma}$ from light-by-light scattering at the LHC in proton-proton~\cite{Baldenegro:2018hng} and $Pb$-$Pb$
collisions~\cite{Knapen:2016moh} (the latter is also significantly weakened in the present scenario by the presence of $g_{agg}$), as they are not competitive with 
the search presented here.

\vspace{-5mm}

\subsection{3) $\mathbf{p p \to a^* \to gg}$ } 
\label{sec:CMS_gg}

\vspace{-3mm}

As discussed previously, the gluonic coupling $g_{agg}$ can be constrained independently, using di-jet searches at the LHC. 
In this work we use the 13 TeV CMS search on di-jet angular distributions~\cite{CMSdijets} for this purpose.  
The details of our selection procedure are given in Appendix A3. 
A 95\% C.L. excluded  region of
 $1.9\,\, \mathrm{TeV} > f_a > 3$ TeV for $c_{\tilde{G}} = 1$ follows from our analysis. The latter 
includes the interference between the ALP signal and the SM background, which dominates over the pure ALP signal in the region considered, as already mentioned. 
 Note that the  limit obtained is weaker than the benchmark value 
$g_{agg}^{-1} = 1$ TeV used in Fig.~\ref{Fig_Bounds} (corresponding to $f_a/c_{\tilde{G}} = 4$ TeV), as it should be. 
This bound is  to be taken only as a qualitative estimate, as the analysis uses data in the 2.4 to 3 TeV range, and thus in the limit of validity of the EFT.

\vspace{-4mm}

\subsection{4) Further non-resonant ALP searches}

\vspace{-3mm}

\noindent {\bf{\it(i)}} $\mathbf{p p \to a^* \to Z h}$. This process yields a powerful probe of the chiral EFT through the 
operator~$\A_{2D}$ in Eq.~\eqref{La_LO}. For $Z \to \ell \ell$ and $h \to \bar{b} b$ this signature 
is similar to that analyzed in Sect. IV-1 for the {\sl $b$-tagged} category, since
the process $p p \to a^* \to Z h$ has similar $m_{\ell\ell J}$ kinematics and expected cross section than  $p p \to a^* \to Z Z$,  for $c_{2D} \simeq c_{\tilde{W}}, c_{\tilde{B}}$.
This suggests that the analysis performed above could be adapted  to probe very efficiently the ALP-mediated $Zh$ signal. 
Furthermore, there are several advantages in performing a dedicated $Zh$ search along the lines in Ref.~\cite{Sirunyan:2018hsl}:  the SM background distribution for 
the merged jet mass $m_J$ is smaller around $m_h$ than around $m_Z$ (as shown in Ref.~\cite{Sirunyan:2018hsl}),  and   
 the SM backgrounds after the CMS event selection are significantly smaller in the {\sl $b$-tagged} category, as 
shown in Fig.~\ref{Fig_DistSignalmZZ}; iii) $h$ decays dominantly to $\bar{b} b$.

\vspace{1mm}

\noindent {\bf{\it(ii)}} $\mathbf{p p \to a^* \to t \bar{t}}$. This channel allows to probe the ALP-fermion couplings induced by the 
operator $\mathcal{O}_{a\Phi}$ in the ALP Linear EFT.   Because the amplitude of any physical ALP-fermion coupling is proportional to the fermion Yukawa couplings  
 (see Eq.~\eqref{eq:leff-alp}  and Footnote 1), ALP production via gluon fusion with $t\bar{t}$ in the final state is an optimal channel which deserves detailed future study.\footnote{ Signals sensitive to the ALP-top coupling  have been briefly discussed in Ref.~\cite{Brivio:2017ije} in the context of the non-linear operator $\A_{2D}(h)$. Nevertheless, the bound obtained  followed from  $aZh$ interactions contained in $\A_{2D}(h)$, which do no apply for $\mathcal{O}_{a\Phi}$ (i.e. in the linear EFT expansion at NLO).  }

\vspace{1mm}  
 
\noindent {\bf{\it(iii)}} $\mathbf{p p \to a^* \to Z \gamma}$. This channel provides a key complementary probe to the $ZZ$ and $\gamma\gamma$ searches discussed in 
Sect.~IV-1) and -2, given its clean signature. The non-resonant analysis of this channel using public information 
(e.g. from Ref.~\cite{Sirunyan:2017hsb}) requires however further assumptions w.r.t. the $ZZ$ and $\gamma\gamma$ analyses. The study of ALP-mediated $Z\gamma$ signatures is thus left for a forthcoming work~\cite{GNST}.

\vspace{-5mm} 
 
%%%%%%%%%%%%%%%%%%%%%%%%%%%%%%%%%%%%%%%%%%%%%%%%%%%%%%%%%%%%
\section{V. Conclusions \& Outlook}
%%%%%%%%%%%%%%%%%%%%%%%%%%%%%%%%%%

\vspace{-3mm}

In this work we have proposed a new approach to probe the existence of ALPs (and more generally of pseudo-Goldstone bosons),   
via non-resonant searches at the LHC where the ALP can be produced as an $s$-channel off-shell mediator.   
The search takes advantage of the derivative nature of the ALP interactions with SM particles.
Using CMS 13 TeV public data, we have derived new limits on ALP couplings to SM gauge bosons 
via the processes $p p \to Z Z$, $p p \to \gamma \gamma$ and $p p \to j j \, (g g)$. These provide   the most stringent bounds
on ALPs over a wide region of masses, in the presence of an ALP-gluonic coupling $g_{agg}$.  They have the advantage of being 
equally sensitive to light ALPs with masses up to the kinematical energy scale of the LHC analyses 
considered $\sim \mathcal{O}(100)\, \mathrm{GeV}$.
Possible extensions of the analysis to other final states such as $Z\gamma$, $Zh$ and fermionic final states ($t \bar{t}$) have been discussed as well, altogether 
highlighting the power of non-resonant searches for ALPs at colliders.

\vspace{2mm}

\begin{center}
\textbf{Acknowledgements} 
\end{center}

\vspace{-1mm}

\begin{acknowledgements}
We thank Gonzalo Alonso-Alvarez, Jes\'us Bonilla, Pablo Quilez and Stefan Pokorsky for useful discussions and comments on the manuscript. 
The work of V.S. is funded by the Science Technology and Facilities Council (STFC) under grant number ST/P000819/1. 
M.B.G.  and J.F.T. acknowledge support  from the ``Spanish Agencia Estatal de Investigaci\'on'' (AEI) and the EU ``Fondo Europeo de Desarrollo Regional'' 
(FEDER) through the projects FPA2016-78645-P and FPA2017-84260-C3-2-R, respectively.
The work of J.M.N. was supported by the Programa Atracci\'on de Talento de la Comunidad de Madrid via 
grant 2017-T1/TIC-5202.
M.B.G. and J.M.N.  acknowledge support from the European Union's Horizon 2020 research and innovation programme under the Marie Sklodowska-Curie grant 
agreements 690575  (RISE InvisiblesPlus) and  674896 (ITN ELUSIVES), as well as from  
the Spanish Research Agency (Agencia Estatal de Investigaci\'on) through the grant IFT Centro de Excelencia Severo Ochoa SEV-2016-0597.
In addition, they warmly thank Venus Keus for hospitality at the University of Helsinki during the very last stages of this work.

\end{acknowledgements}

\appendix

\vspace{-3mm}

%%%%%%%%%%%%%%%%%%%%%%%%%%%%%%%%%%%%%%%%%%%%%%%%%%%%%%%%%%%%
\section{Appendix A: Details on LHC analyses}
%%%%%%%%%%%%%%%%%%%%%%%%%%%%%%%%%%

\vspace{-3mm}

\begin{figure}[h]
\begin{center}
\includegraphics[width=0.48\textwidth]{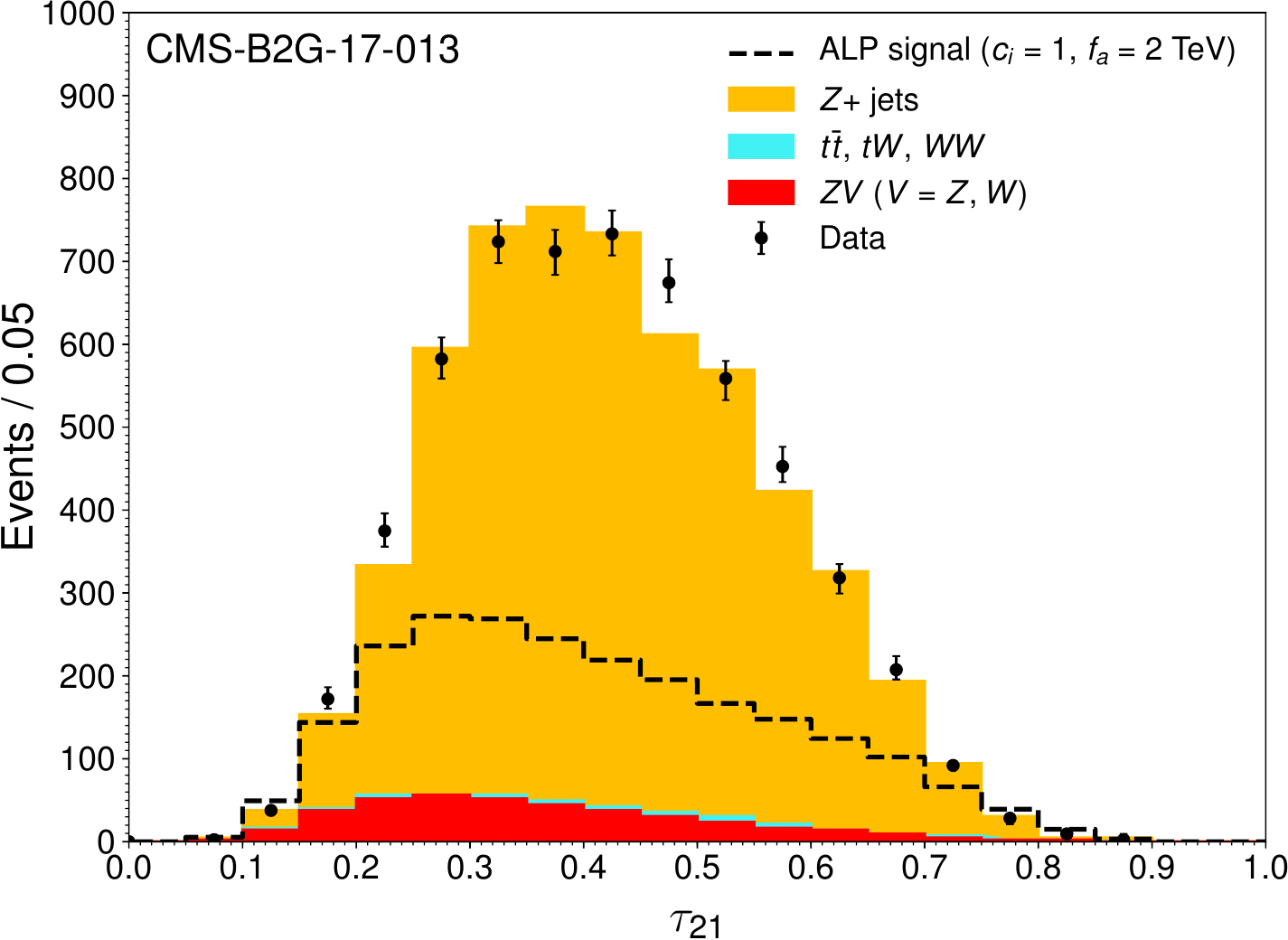}
\caption{\small $\tau_{21}$ differential distribution for the ALP $ZZ$ signal obtained with {\tt Delphes 3} (dashed black line), together with 
the main SM backgrounds and experimental data from~\cite{Sirunyan:2018hsl}, after CMS event selection (without 
the $\tau_{21} < 0.4$ cut).}
\label{Fig_tau21_ZZ_CMS}
\end{center}
\end{figure}

\noindent $\mathbf{1.\,\,CMS\,\, ZZ}$~\cite{Sirunyan:2018hsl}. The analysis requires the leading (subleading) lepton from the event 
to have $p_T > 40$ ($30$) GeV and $|\eta| < 2.1$ ($2.4$), and the invariant mass of the di-lepton pair is required to fall within $70$ GeV $< m_{\ell\ell} < 110$ GeV 
and have $p^{\ell\ell}_{T} > 200$ GeV.
In addition, the ``low-mass merged" category contains an anti-$k_T$ jet with a large radius $R = 0.8$ (AK8) and $p_{T}^J > 200$ GeV. 
The merged jet mass is required to be in the range $65$ GeV $< m_{J} < 105$ GeV.
The analysis further makes use of the information on the subjettiness variables $\tau_{1}$ and $\tau_2$ for the reconstructed AK8 jet to 
build $\tau_{21} = \tau_2/\tau_1$~\cite{Thaler:2010tr}.
The $\tau_{21}$ distribution for the signal (from {\tt Delphes 3}), SM backgrounds and experimental data (obtained from~\cite{Sirunyan:2018hsl}) 
which pass the above selection criteria is shown in Fig.~\ref{Fig_tau21_ZZ_CMS} for the combination of  {\sl untagged} and {\sl $b$-tagged} 
selection categories, and the CMS event selection then requires $\tau_{21} < 0.4$ for the AK8 merged jet.
Overall, the CMS analysis translates into an average ALP signal selection efficiency of $\sim 8\%$.

\vspace{2mm}

\begin{figure}[h]
\begin{center}

\includegraphics[width=0.48\textwidth]{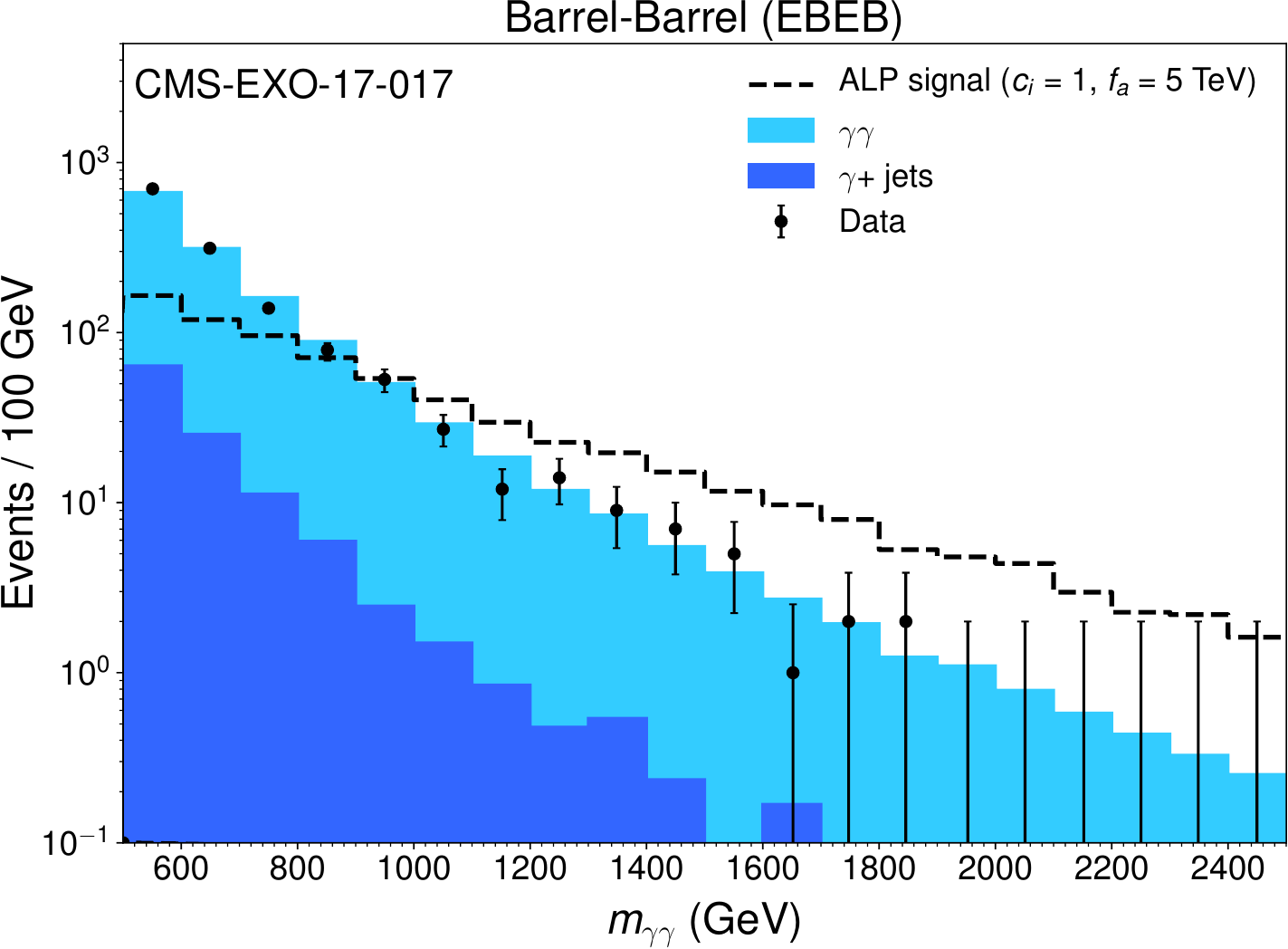}

\vspace{1mm}

\includegraphics[width=0.48\textwidth]{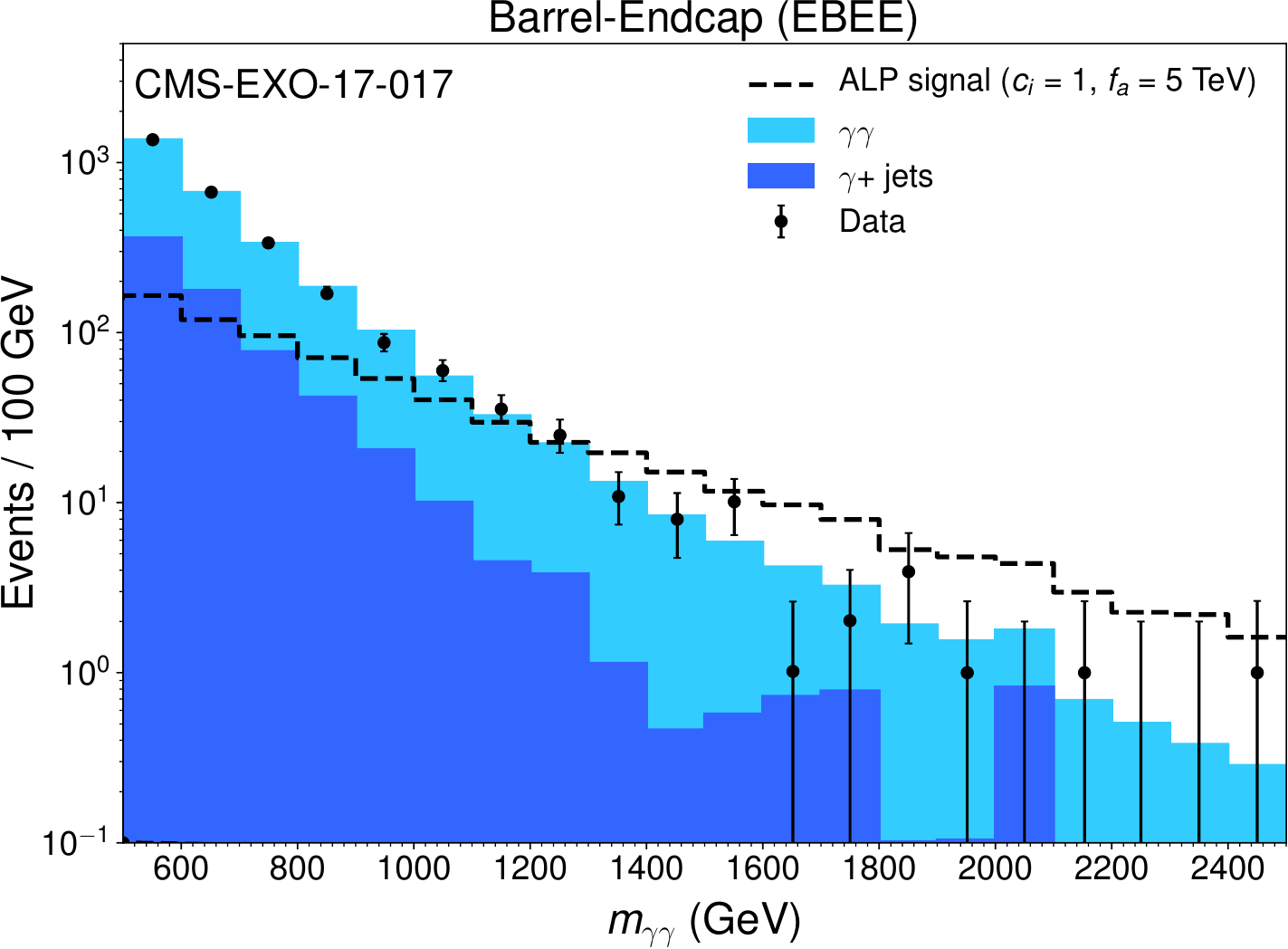}

\vspace{-2mm}

\caption{\small $m_{\gamma\gamma}$ distributions for the ALP $\gamma\gamma$ signal with 
$c_i = 1$, $f_a = 5$ TeV (dashed black line) and SM background from $\gamma\gamma$ (light blue) and $\gamma +$ jets (dark blue)
 after CMS event selection, in the EBEB (top) and EBEE (bottom) categories. The experimental data are shown as black dots.}
\label{Fig_DistSignalmGammaGamma}
\end{center}
\end{figure}

\noindent $\mathbf{2.\,\,CMS\,\, \gamma\gamma}$~\cite{Sirunyan:2018wnk}. The analysis requires two photons with $p_T > 75$ GeV each, 
an invariant mass $m_{\gamma\gamma} > 500$ GeV and a distance $\Delta R_{\gamma\gamma} > 0.45$. One of the photons has to be detected in 
the electromagnetic calorimeter (ECAL) barrel (EB) region, corresponding to $\left|\eta \right| < 1.44$. 
The other photon can either be detected in the EB region or in the ECAL endcap (EE) region, $1.57 < \left|\eta \right| < 2.5$, 
respectively defining two distinct analysis regions (labelled EBEB and EBEE) for the search.
The CMS reconstruction efficiency for EB (EE) photons in the signal region is approximately $0.90$ ($0.87$)~\cite{Sirunyan:2018wnk}. 
We find the CMS analysis yields an average ALP signal selection efficiency of $\sim 72\%$ for a signal sample with $m_{\gamma\gamma} > 500$ GeV.
The di-photon invariant mass $m_{\gamma\gamma}$ distribution after event selection for the ALP signal (with $c_{\tilde G} = c_{\tilde W} = c_{\tilde B} = 1$ and 
$f_a = 5$ TeV) and SM backgrounds is shown in Fig.~\ref{Fig_DistSignalmGammaGamma} for the EBEB (top) and EBEE (bottom) categories. 
Combining both EBEB and EBEE categories we obtain an observed 95\% C.L. exclusion limit of $f_a > 14.2$ TeV for $c_i = 1$, quoted in 
Sect. 4-2. 

\vspace{2mm}

\begin{figure}[t]
\begin{center}
\includegraphics[width=0.48\textwidth]{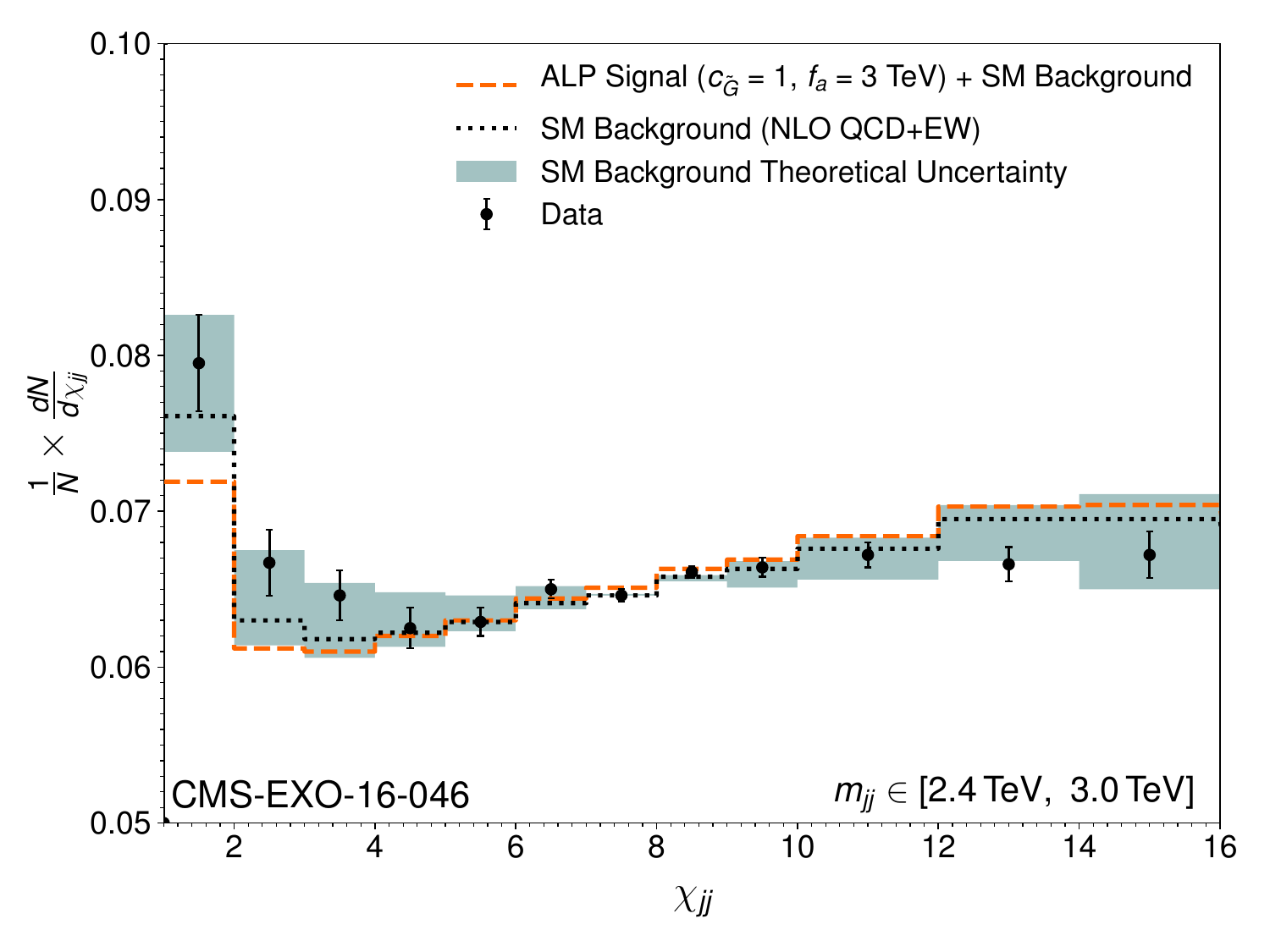}
\caption{\small Di-jet differential distribution as a function of the angular variable $\chi_{jj}$ in the bin with di-jet invariant masses 2.4 TeV - 3.0 TeV. 
The dotted black line corresponds to the QCD SM background (with its theoretical uncertainty shown as a grey band). 
 The dashed red line corresponds to a normalized combination of the ALP signal for $f_a/c_{\tilde G} = 3 $ TeV and the QCD background.
The experimental data from the CMS di-jet analysis~\cite{CMSdijets} are shown in black. 
The SM background and experimental data have been taken from HEPDATA~\cite{hepdata}.}
\label{Fig_dijet_CMS}
\end{center}

\vspace{-7mm}

\end{figure}

\noindent $\mathbf{3.\,\,CMS}$ {\bf di-jet}~\cite{CMSdijets}. The analysis selects two $R = 0.4$ anti-$k_T$ (AK4) jets 
with $|\eta_j|< $ 2.5, $p_T^j> $ 200 GeV, 
$m_{jj} > 2.4$ TeV and also imposes angular cuts on the di-jet system, $y_{\textrm{boost}} = (y^j_{1} + y^j_{2})/2 < 1.11$ (where  
$y^j_{{1,2}}$ denote the rapidities of the two jets) and $\chi_{jj} = \mathrm{exp} (|y^j_{1} - y^j_{2}|) < 16$.
 The analysis is restricted in this work to the first invariant mass bin considered in Ref.~\cite{CMSdijets}, 
$m_{jj} \in [$2.4, 3.0] TeV. After event selection, the CMS analysis provides the normalized $\chi_{jj}$ distribution for the experimental data 
and for the SM background prediction. 
We computed the normalized $\chi_{jj}$ signal distribution after CMS events selection using {\tt MadGraph$\_$aMC@NLO}, 
{\tt Pythia 8} and {\tt FastJet}~\cite{Cacciari:2011ma}, including both the pure ALP signal, scaling as $(c_{\tilde{G}}/f_a)^{4}$, and 
the interference with the SM background, scaling as $(c_{\tilde{G}}/f_a)^{2}$. The size of the pure ALP signal and that of the
(negative) interference with the QCD background are approximately equal for $f_a/c_{\tilde{G}} \sim 1.5$ TeV, and the 
interference dominates for larger values of $f_a$ (those relevant for our analysis, since we look at invariant masses $m_{jj} \in [$2.4, 3.0] TeV).  
 A $\chi^2$ fit to the data was performed using a linear combination of the normalized QCD background and ALP signal, 
with relative weights \ $(1-q)$ and $q$ respectively. Our procedure results in a 95\% C.L. excluded region of $1.9 \,\,\mathrm{TeV} > f_a > 3$ 
TeV for $c_{\tilde G} = 1$ (corresponding to an excluded ALP signal weight $q <-0.016$). Fig.~\ref{Fig_dijet_CMS} shows the normalized $\chi_{jj}$ distribution in 
 the invariant mass bin $m_{jj} \in [$2.4, 3.0] TeV, after the CMS event selection for the QCD background (with its  theoretical uncertainty), as well as a normalized  
 combination of the ALP signal for $f_a/c_{\tilde G} = 3 $ TeV and the QCD background.

%%%%%%%%%%%%%%%%%%%%%%%%%%%%%%%%%%%%%%%%%%%%%%%%%%%%%%%%%%%%%%%

\end{document}